\documentstyle[aps,prl]{revtex}
\begin{document}
\draft
\twocolumn[
\title{Simple Maps with Fractal Diffusion Coefficients}
\author{R. Klages\cite{em}}
\address{Institut f\"ur Theoretische Physik, Technische Universit\"at
Berlin, Sekr.\ PN 7-1,\\ Hardenbergstr.\ 36, D-10623 Berlin, Germany}
\author{J.R. Dorfman}
\address{Institute for Physical Science and Technology and
Department of Physics,\\ University of Maryland, College Park, MD
20742, USA}
\date{\today}
\maketitle
\begin{abstract}
\widetext
\vspace{-0.5cm}
We consider chains of one-dimensional, piecewise linear, chaotic maps
with uniform slope. We study the diffusive behaviour of an initially
nonuniform distribution of points as a function of the slope of the
map by solving the Frobenius-Perron equation.  For Markov partition
values of the slope, we relate the diffusion coefficient to
eigenvalues of the topological transition matrix. The diffusion
coefficient obtained shows a fractal structure as a function of the
slope of the map. This result may be typical for a wide class of maps,
such as two-dimensional sawtooth maps.\\
\narrowtext
\end{abstract}
\pacs{PACS numbers: 05.45.+b, 05.60.+w}]
The study of simple models for non-equilibrium processes in
statistical physics has been one of the central themes in the theory
of chaotic dynamical systems \cite{swl,wig,ott}. A great deal of work
has been done to describe the large-scale motion in systems of
independent particles each moving under the action of relatively
simple maps, operating at discrete intervals of time. For
one-dimensional sinusoidal, or piecewise differentiable maps, a variety of
diffusive-like and ballistic-like behaviour has been studied
\cite{gro,sfk,gg,art}. For two-dimensional, conservative Hamiltonian
maps, parameter dependent momentum-diffusion coefficients have been
computed, often by a combination of numerical and analytical methods
which explore the phase space structure of the dynamical system
\cite{rw,cm,da}. Recently, Gaspard and coworkers have established an
explicit connection between fundamental quantities of dynamical
systems, such as the Kolmogorov-Sinai entropy and Lyapunov exponents,
and transport coefficients \cite{ga1,jrd,ga2,ga3}. Related connections
between transport coefficients and Lyapunov exponents have been
discussed for non-equilibrium systems with thermostats \cite{ech}.
There is also a close connection of the work described here to that
based on periodic orbit expansions for transport coefficients
\cite{art,cvi}.

In this letter, chains of piecewise linear, one-dimensional, chaotic maps
\begin{equation}
x_{\tau+1} = [x_{\tau}] + m_a(x_{\tau}) \equiv M_a(x_{\tau}) \label{ds}
\end{equation}
will be considered, where $\tau$ is the discrete time variable, $
[x_{\tau}] $ the largest integer smaller than $x_{\tau}$,
$m_a(x_{\tau}+1) = m_a(x_{\tau})$ represents a periodic function, and
$a$ stands for the control parameter, which is the slope of the map.
We consider a chain of maps $m_a(x_{\tau})$ with chainlength $L$. The
absolute value of the slope is assumed to be uniform and its logarithm
is equal to the Lyapunov exponent. We assume the maps are expanding,
i.e.\ that $|a| > 1$.  Following the approach in
\cite{ga1,jrd,ga2,ga3}, we describe a method by which the diffusion
coefficient for this class of maps can be computed for a broad range
of parameter values. The method will be illustrated by the map
\begin{equation}
m_a(x_{\tau}) := \left\{ \begin{array}{r@{\quad , \quad}l}
a x_{\tau} & 0 < x_{\tau} \leq \frac{1}{2} \\
a x_{\tau}+1-a & \frac{1}{2} < x_{\tau} \leq 1 \end{array} \right.
\quad , \quad a>0 \quad ,
\label{map}
\end{equation}
as sketched in Fig.~\ref{f1}, which has been introduced and discussed
in \cite{ott,gro,ga2}. We find that the diffusion coefficient for this
map shows a very rich fractal structure as a function of the slope.

To describe the dynamical behaviour of an arbitrary initial density
for a set of particles on some interval of the line $-\infty \leq x
\leq \infty$, we will need the Frobenius-Perron equation, supplemented
by boundary conditions. The Frobenius-Perron equation is given by
\begin{equation}
\rho_{\tau+1}(x) = \int dy \; \rho_{\tau}(y) \; \delta(x-M_a(y)) \quad ,
\label{fp}
\end{equation}
where $\rho_{\tau}(x)$ is the probability density for points on the
line, and $M_a(y)$ is the map under consideration.  We suppose that
the motion takes place on an interval $0<x<L$, and we impose periodic
boundary conditions, i.e.\ $\rho_{\tau}(0)=\rho_{\tau}(L)$ for all
$\tau$, or absorbing boundary conditions $\rho_{\tau}(x)=0$ for
$x=0,L$ for all $\tau$ \cite{vk}.  Next we use the argument of Gaspard
and coworkers \cite{ga1,jrd,ga2,ga3} to relate the eigenmodes of the
Frobenius-Perron equation to the solution of the diffusion equation
\begin{equation}
\frac{\partial n(x,t)}{\partial t} = D \;
\frac{\partial^2n(x,t)}{\partial x^2} \quad ,
\end{equation}
where $n(x,t)$ is the macroscopic density of particles at a point $x$
at time $t$, and $D$ is a diffusion coefficient. If for large $L$, and
large $\tau$, the first few eigenmodes of the Frobenius-Perron
equation are identical to those of the diffusion equation, the
diffusion coefficient can be obtained by matching eigenmodes in an
appropriate scaling limit. More explicitly, for periodic boundary
conditions $n(0,t)=n(L,t)$, one can easily see that for large times
$n(x,t)$ has the form
\begin{equation}
n(x,t) = const. +A\exp\left(-D(4\pi^2/L^2)t\pm
i(2\pi/L)x\right) \quad .
\end{equation}
Consequently, if one can find a solution of Eq.~(\ref{fp}), for large
$L$ and $\tau$, in the form of
\begin{equation}
\rho(x,\tau)=const. +A^{\prime}\exp\left(-\gamma_p(a)\tau\pm
i(2\pi/L)x\right) \quad ,
\end{equation}
one can relate the decay rate $\gamma_p(a)$ to $D$ by
\begin{equation}
D(a) = \lim_{L\to\infty} \left(L/2\pi\right)^2\;
\gamma_p(a) \quad . \label{gdec}
\end{equation}
For absorbing boundary conditions one relates the diffusion
coefficient to the escape rate from the system by an equation similar
to Eq.(\ref{gdec}). The escape-rate formalism of chaotic dynamics
shows that the escape rate from a system with absorbing boundaries is
equal to the Lyapunov exponent minus the Kolmogorov-Sinai entropy for
particles trapped within the system whose trajectories lie on a
fractal repeller \cite{ga1}. For the case of maps with slopes of
uniform magnitude considered here, the Komolgorov-Sinai entropy for
the fractal repeller $h_{KS}(a)$ is identical to the topological
entropy of points on the repeller and it can be computed from
$\gamma_p(a)$ as $h_{KS}(a)=\log a - \gamma_p/4+O(L^{-3}) \:$
\cite{ga2}.

The use of maps with uniform slope is not an essential ingredient in
the calculation of $D(a)$ described below, which can be applied to
more general linear maps.  The main idea is that the Frobenius-Perron
equation can be written as a matrix equation whenever the parameters
of the map are such that one can construct a Markov partition of the
interval $(0,L)$, which has the property that partition points get
mapped onto other partition points by the map $M_a(x)$ \cite{boy}. In
a related context, these partitions have been discussed in \cite{bst}.
For such values of $a$, Eq.(\ref{fp}) can be written as
\begin{equation}
\mbox{\boldmath $\rho$}_{\tau+1}=(1/|a|) \, M \, \mbox{\boldmath $\rho$}_
{\tau} \quad ,
\end{equation}
where $\mbox{\boldmath $\rho$}_{\tau}$ is a column vector of the
probability densities in each of the Markov partition regions at time
$\tau$, and $M$ is a topological transition matrix whose elements
$M_{ij}$ are unity if points in region $j$ can be mapped into region
$i$, and are zero otherwise.

As a simple example we consider the form of the matrix $M$ when $a=3$,
the map $M_a(x)$ is given by Eqs.(\ref{ds}),(\ref{map}), and periodic
boundary conditions are used on an interval of length $L$. In this
case the regions of the partition are all of length $1/2$, as
illustrated in Fig.~\ref{f1}. Then $M$ is a $2L$ x $2L$ matrix of the
form
\begin{equation}
M=
\left( \begin{array}{ccccccccc} 1 & 1 & 0 & 0 & \cdots &0 & 0 & 1 & 0 \\
1 & 1 & 0 & 1 & 0 & 0 & \cdots & 0 & 0 \\
1 & 0 & 1 & 1 & 0 & 0 & \cdots & 0 & 0 \\
0 & 0 & 1 & 1 & 0 & 1 & 0 & 0 & \cdots \\
0 & 0 & 1 & 0 & 1 & 1 & 0 & 0 & \cdots \\
\vdots & & & & \vdots & & & & \vdots \\
0 & 1 & 0 & 0 & \cdots & 0 & 0 & 1 & 1 \\
\end{array} \right) \quad . \label{tm}
\end{equation}
In the limit $\tau\to\infty$, for any $L$, and any ``Markov partition''
value of $a$, the Frobenius-Perron equation can be solved in terms of
the eigenmodes of $M$ for any initial value $\rho_0(x)$ which is
uniform in each of the Markov partition regions. For periodic boundary
conditions, $M$ is always a (block)circulant \cite{bk}, the largest
eigenvalue of $M$ is precisely $|a|$, and the corresponding eigenmode
is a constant, representing the equilibrium state. The rate of decay
to equilibrium, $\gamma_p(a)$, is obtained as
$\gamma_p(a)=\log\left(|a|/\chi_1\right)$, where $\chi_1$ is the next
largest eigenvalue of $M$ \cite{ga2}.  Analytical expressions for
$D(a)$ can be derived for all integer values of $a\geq2$. For even
integers, the results of Grossmann and Fujisaka \cite{gro} are
recovered, $D(a)=(1/24)(a-1)(a-2)$, and for odd integers we find
$D(a)=(1/24)\left(a^2-1\right)$. To obtain $D(a)$ for a general Markov
partition value of $a$, one can use computer methods \cite{evpnu}.

Fig.~\ref{f2} (a) shows the results for the diffusion coefficient of
the dynamical system Eqs.(\ref{ds}),(\ref{map}) for values of $a$ in
the range $2\leq a \leq 8$. In Fig.~\ref{f2} (b)-(d), we present
magnifications of three small regions in this interval \cite{nup}. One
can see clearly that $D(a)$ has a complicated fractal structure with
regions exhibiting self-similarity. In Fig.~\ref{f3}, we show an
enlargement of the region for $2\leq a \leq 3$. The dashed line is the
prediciton of $D(a)$ for a simple random-walk model suggested by
Schell, Fraser and Kapral \cite{sfk}. Note that the model correctly
accounts for the behaviour of $D(a)$ near $a=2$. The wiggles in this
graph can be understood by considering the transport of particles from
one unit interval to another. These regions are coupled to each other
by {\em turnstiles}, where points in one unit interval get mapped
outside that particular interval into another unit interval. As in the
case of two-dimensional twist maps, such as the sawtooth map, these
turnstiles are crucial for large-scale transport \cite{wig,mch}.

The region $2\leq a \leq 3$ can be analyzed by studying the
interaction of turnstiles \cite{rk}. One can recognize three distinct
series of values of $a$, each of which provides a cascade of
apparently self-similar regions of decreasing size, as the limits $a
\rightarrow 2$ or $a \rightarrow 3$ are approached. To understand
these series, consider the trajectory of a point that starts just to
the left at $x=1/2$. The first iterate of $x=1/2$ is in the second
interval, $(1,2)$. The {\em series $\alpha$} values of $a$ are defined
by the condition that the second iterate of $x=1/2$ is at the leftmost
point of the upward turnstile in the second interval $(1,2)$
$(a=2.732)$, or that the third iterate is at the corresponding point
in the third interval $(a=2.920)$, etc.\ The numbers on the graph
refer to the number of intervals the image of $x=1/2$ has travelled
before it gets to the appropriate point on the turnstiles. {\em Series
$\beta$} points are defined in a similar way, but they are allowed to
have two or more internal reflections within an interval before
reaching the left edge of a turnstile. {\em Series $\gamma$} points
are defined by the condition that some image of $x=1/2$ has reached
the rightmost edge of an upward turnstile (i.e.\ some point
$x=n+1/2$, where $n$ is an integer), and consequently an increase in
$a$ will lead to a decrease in $D(a)$. These cascades provide a basis
for a physical understanding of the features of $D(a)$ in this region:
Particles leave a particular unit interval through a turnstile and
undergo a number of iterations before they are within another
turnstile. Whether they continue to move in the same or the reverse
direction at the next and later turnstiles is a sensitive function of
the slope of the map. Thus the fractal structure of the $D(a)$ curve
is due to the effects of long-range correlations among turnstiles and
these correlations lead to changes of $D(a)$ on an infinitely fine
scale. A similar argument can be employed to explain, at least
qualitatively, the fractal structure of $D(a)$ for higher values of
the slope, although more work needs to be done before a full
understanding of this curve is obtained \cite{tsd}.

We conclude with a few remarks:
(A) Our results appear to be the first example of a system whose
diffusion coefficient has an unambiguously fractal structure. We
suspect that similar results obtain for all other one-dimensional,
piecewise linear maps \cite{tsg}, which might be of interest, e.g.,
for chaotic scattering \cite{lg}, as well as for transport in maps of
more than one dimension, such as sawtooth maps. We note that
oscillations of the diffusion coefficient with respect to an
appropriate control parameter, which could be a field strength, have
already been found in the standard \cite{rw} and the sawtooth map
\cite{da}.
(B) It is not known whether this fractal structure persists for smooth
maps where the function $M_a(x)$ is $C^1$, or where the map contains
some randomness.
(C) We have numerical evidence that the Markov points are dense for $a
\geq 2$, and we believe that our results give the full structure of
the $D(a)$ function. Nevertheless, it would be valuable to have a
mathematical proof.

We are indebted to J.\ Yorke for an important hint regarding Markov
partitions. Helpful discussions with Chr.\ Beck, C.\ Grebogi, R.\
Kapral, L.\ Bunimovich, and B.\ Hunt are gratefully acknowledged.
R.K.\ wants to thank the Institute for Physical Science and Technology
for its hospitality, and he is grateful to S.\ Hess, T.R.\
Kirkpatrick, the DAAD and the NaF\"oG commission Berlin for financial
and other support.

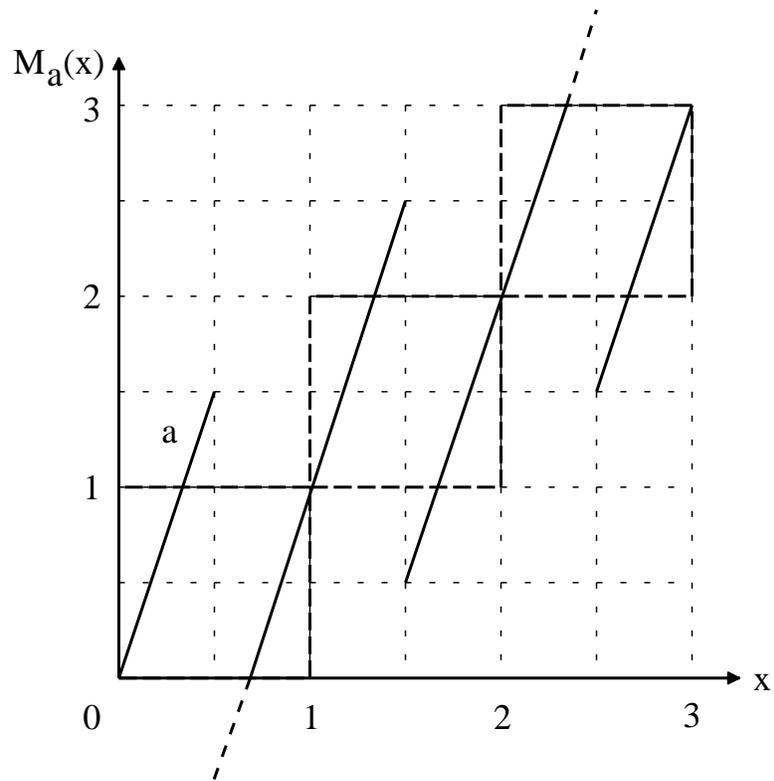
\begin{figure}
\caption{Illustration of the dynamical system
Eqs.~(\protect\ref{ds}),(\protect\ref{map}) for a particular slope, $a=3$.
The Markov partition given by the dashed grid leads to the
construction of the transition matrix in Eq.~(\protect\ref{tm}). }
\label{f1}
\end{figure}
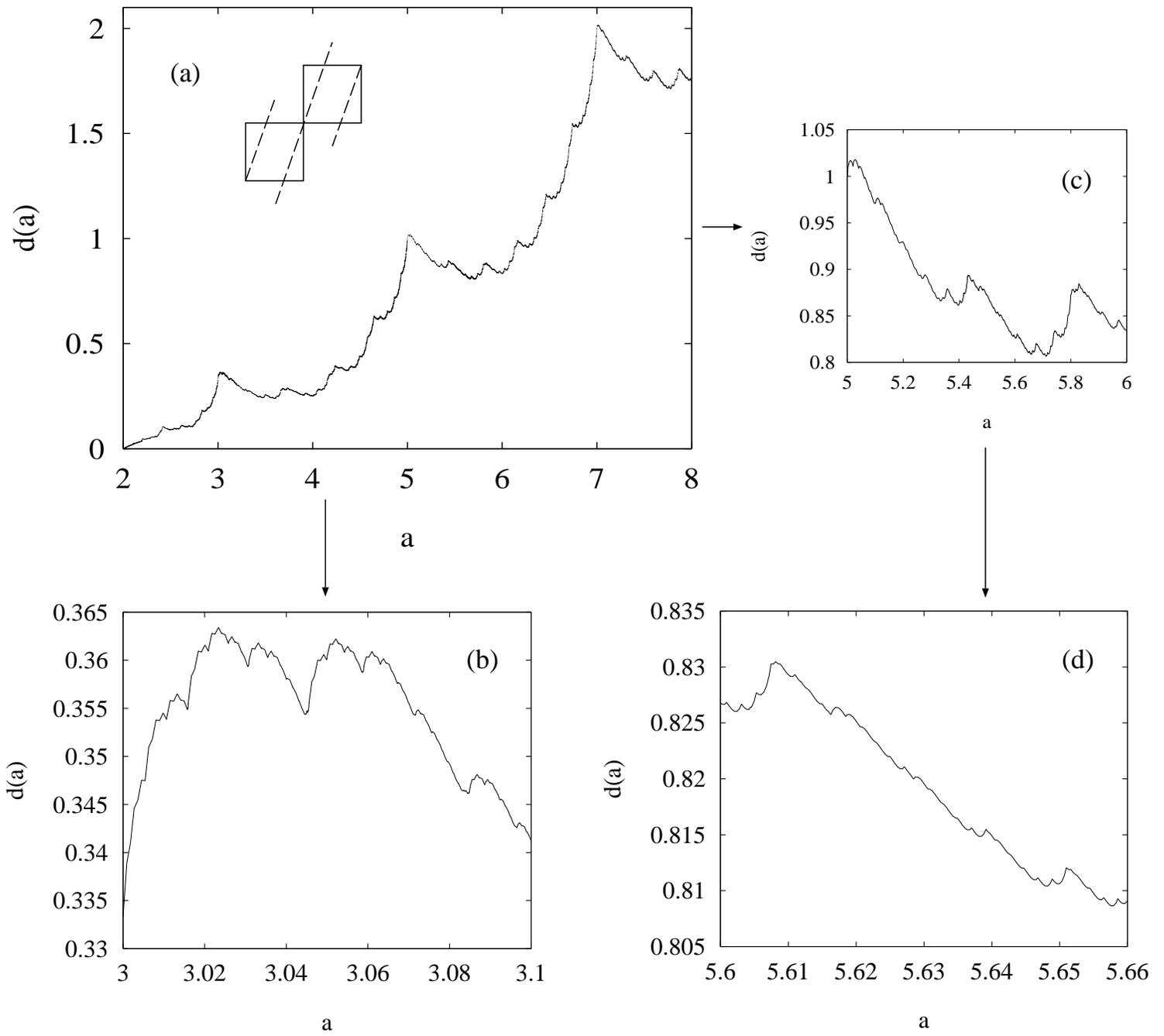
\begin{figure}
\caption{Diffusion coefficient $D(a)$ computed for the dynamical
system Eqs.~(\protect\ref{ds}),(\protect\ref{map}) and some enlargements.
Graph (a) consists of 7908 single data points. In graph (b)-(d), the
dots are connected with lines. The number of data points is 476 for
(b), 1674 for (c), and 530 for (d).}
\label{f2}
\end{figure}
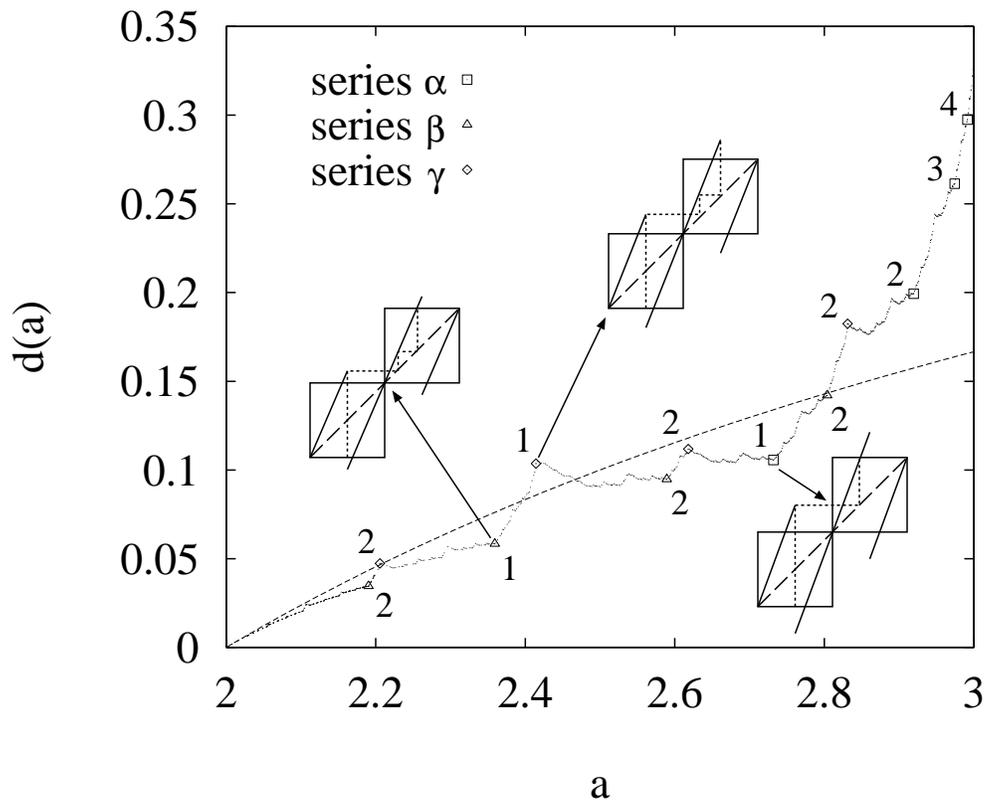
\begin{figure}
\caption{Enlargement of the region of slope $a \leq 3$ with the solution for
a simple random walk model (dashed line) and labels for the values
which are significant for ``turnstile dynamics'' (see text). For
some points, the turnstile coupling is shown by pairs of boxes. The
graph shows 979 single data points.}
\label{f3}
\end{figure}

\begin{references}
\bibitem[*]{em}e-mail: rkla0433@w421zrz.physik.tu-berlin.de
\bibitem{swl}H.G.\ Schuster, {\em Deterministic Chaos} (VCH
Verlagsgesellschaft mbH, Weinheim, 1989), 2nd edition;
A.J.\ Lichtenberg, M.A.\ Lieberman, {\em Regular and Chaotic
Dynamics.} (Springer, New York, 1983) 2nd edition; M.\ Mareschal, B.L.\
Holian eds., {\em Microscopic Simulations of Complex Hydrodynamic
Phenomena} (Plenum, New York, 1992)
\bibitem{wig}S.\ Wiggins, {\em Chaotic Transport in Dynamical
Systems} (Springer-Verlag, New York, 1992)
\bibitem{ott}E.\ Ott, {\em Chaos in Dynamical Systems} (Cambridge,
1993)
\bibitem{gro}S.\ Grossmann, H.\ Fujisaka, Phys.Rev.\ A {\bf 26}, 1179
(1982); H.\ Fujisaka, S.\ Grossmann, Z.Phys.\ B - Condensed Matter
{\bf 48}, 261 (1982)
\bibitem{sfk}M.\ Schell, S.\ Fraser, R.\ Kapral, Phys.Rev.\ A {\bf
26}, 504 (1982)
\bibitem{gg}T.\ Geisel, J.\ Nierwetberg, Phys.Rev.Lett.\ {\bf 48}, 7
(1982); S.\ Grossmann, S.\ Thomae, Phys.Lett.\ {\bf 97A}, 263 (1983);
T.\ Geisel, S.\ Thomae, Phys.Rev.Lett.\ {\bf 52}, 1936 (1984);
T.\ Geisel, J.\ Nierwetberg, A.\ Zacherl, Phys.Rev.Lett.\ {\bf 54},
616 (1985)
\bibitem{art}R.\ Artuso, Phys.Lett.\ A {\bf 160}, 528 (1991); R.\
Artuso, G.\ Casati, R.\ Lombardi, Phys.Rev.Lett.\ {\bf 71}, 62 (1993)
\bibitem{rw}A.B.\ Rechester, R.B.\ White, Phys.Rev.Lett.\ {\bf 44},
1586 (1980); A.B.\ Rechester, M.N.\ Rosenbluth, R.B.\ White,
Phys.Rev.\ A {\bf 23}, 2664 (1981)
\bibitem{cm}J.R.\ Cary, J.D.\ Meiss, A.\ Bhattacharjee, Phys.Rev.\ A
{\bf 23}, 2744 (1981); J.R.\ Cary, J.D.\ Meiss, Phys.Rev.\ A {\bf 24},
2664 (1981); T.M.\ Antonsen, E.\ Ott, Phys.Fluids {\bf 24}, 1635
(1981)
\bibitem{da}I.\ Dana, N.W.\ Murray, I.C.\ Percival, Phys.Rev.Lett.\
{\bf 62}, 233 (1989); I.\ Dana, Physica D {\bf 39}, 205 (1989)
\bibitem{ga1}P.\ Gaspard, G.\ Nicolis, Phys.Rev.Lett.\ {\bf 65}, 1693
(1990)
\bibitem{jrd}J.R.\ Dorfman, P.\ Gaspard, submitted to Phys.Rev. E
\bibitem{ga2}P.\ Gaspard, J.Stat.Phys.\ {\bf 68}, 673 (1992);
Phys.Lett.\ A {\bf 168}, 13 (1992); Chaos {\bf 3}, 427 (1993)
\bibitem{ga3}P.\ Gaspard, F.\ Baras, in {\em Microscopic Simulations
of Complex Hydrodynamic Phenomena.} op.cit.\
\bibitem{ech}D.J.\ Evans, E.G.D.\ Cohen, G.P.\ Morris, Phys.Rev.\ A
{\bf 42}, 5990 (1990); A.\ Baranyai, D.J.\ Evans, E.G.D.\ Cohen, J.Stat.Phys.\
{\bf 70}, 1085 (1993); N.I.\ Chernov {\em et al.}, Phys.Rev.Lett.\
{\bf 70}, 2209 (1993); Comm.Math.Phys.\ {\em 154}, 569 (1993); H.A.\
Posch, W.G.\ Hoover, Phys.Lett.\ A {\bf 123}, 227 (1987); Phys.Rev.\ A
{\em 39}, 2175 (1989)
\bibitem{cvi}P.\ Cvitanovi\'c, J.-P.\ Eckmann, P.\ Gaspard, Niels Bohr
Institute preprint (May 1991); P.\ Cvitanovi\'c, P.\ Gaspard, T.\
Schreiber, Chaos {\bf 2}, 85 (1992), and references therein
\bibitem{vk}N.G.v.\ Kampen, {\em Stochastic Processes in Physics and
Chemistry} (North Holland, Amsterdam, 1981)
\bibitem{boy}see, e.g., A.\ Boyarski, M.\ Skarowsky, Trans.Am.Math. Soc.\
{\bf 225}, 243 (1979); A.\ Boyarski, J.Stat.Phys.\ {\bf 50}, 213 (1988);
Chr.\ Beck, F.\ Schl\"ogl, {\em Thermodynamics of Chaotic Systems}
(Cambridge, 1993)
\bibitem{bst}C.S.\ Hsu, M.C.\ Kim, Phys.Rev.\ A {\bf 31}, 3253 (1985);
N.\ Balmforth, E.A.\ Spiegel, C.\ Tresser, Phys.Rev.Lett.\ {\bf 72},
80 (1994)
\bibitem{bk}see, e.g., T.H.\ Berlin, M.\ Kac, Phys.Rev.\ {\bf
86}, 8211 (1952); Ph.J.\ Davis, {\em Circulant Matrices} (Wiley, New
York, 1979)
\bibitem{evpnu}For circulant matrices, standard software packages
(NAG, IMSL) do not always give the full spectrum correctly (cf.\ R.M.\
Beam, R.F.\ Warming, NASA Technical Memorandum 103900 (Moffett Field,
California, 1991), unpublished), but the results for the first two
largest eigenvalues are reliable.
\bibitem{nup}For a chainlength of $L=100$, the numerical precision for
each $D(a)$ is always better than 0.1\% with respect to the limit in
Eq.(\ref{gdec}). Therefore, errorbars do not appear in the diagrams.
\bibitem{mch}R.S.\ Mackay, J.D.\ Meiss, I.C.\ Percival, Physica D {\bf
13}, 55 (1984), Q.\ Chen, J.D.\ Meiss, Nonlinearity {\bf 39}, 347
(1989); Q.\ Chen {\em et al.}, Physica D {\bf 46}, 217 (1990); J.D.\ Meiss,
Rev.Mod.Phys.\ {\bf 64}, 795 (1992)
\bibitem{rk}R.\ Klages, J.R.\ Dorfman, in preparation
\bibitem{tsd}These qualitative explanations are not sufficient to get
the exact values for all the local extrema of $D(a)$. Compare, e.g.,
$a=3,5,7,\ldots$
\bibitem{tsg}After this letter was submitted we were informed that
related results, based on the methods of Ref.\cite{cvi}, have been
obtained for another one-dimensional map by H.-C.\ Tseng et al.,
submitted to Phys.\ Lett.\ {\bf A}.
\bibitem{lg}Y.-Ch.\ Lai, C.\ Grebogi, Phys.Rev.\ E {\bf 49}, 3761
(1994); Y.-Ch.\ Lai {\em et al.}, Phys.Rev.Lett.\ {\bf 71}, 2212 (1993)
\end{references}
\end{document}